\documentclass[sort&compress,numbers]{aipproc}
\usepackage{graphicx}
\usepackage{epsfig}
\usepackage{amsmath}
\usepackage{appendix}

\layoutstyle{6x9}

\def\lsim{\mathrel{\hbox{\rlap{\hbox{\lower4pt\hbox{$\sim$}}}\hbox{$<$}}}}

\makeatletter

\newcommand{\Rmnum}[1]{\expandafter\@slowromancap\romannumeral #1@}
\makeatother

\begin{document}

\title{Channeling Effects in Direct Dark Matter Detectors\footnote{To appear in the Proceedings of the sixth International Workshop on the Dark Side of the Universe (DSU2010), Leon, Guanajuato, Mexico 1-6 June 2010.}}

\author{Nassim Bozorgnia}{
address={Department of Physics and Astronomy, UCLA, 475 Portola Plaza, Los
  Angeles, CA 90095, USA},
email={nassim@physics.ucla.edu}}

\begin{abstract}
\noindent
The channeling of the ion recoiling after a collision with a WIMP changes the ionization signal
in direct detection experiments, producing a larger signal than otherwise expected. We
give estimates of the fraction of channeled recoiling ions in NaI (Tl), Si and Ge crystals
using analytic models produced since  the 1960's and 70's  to describe channeling and blocking effects.
We find that the channeling fraction of recoiling lattice nuclei is smaller than that of ions  that are injected into the  crystal and that it is strongly temperature dependent.
\end{abstract}



\maketitle
\section{Introduction}

Channeling and blocking effects in crystals refer to the orientation dependence of charged ion penetration in crystals. In the ``channeling effect" ions incident upon a crystal along symmetry axes and planes suffer a series of small-angle scatterings that maintain them in the open ``channels"  between the rows or planes of lattice atoms and thus penetrate much further into the crystal than in other directions. Channeled incoming ions do not get close to lattice sites, where they would be deflected at large angles. The ``blocking effect"  consists in a reduction of the flux of ions originating in lattice sites along symmetry axes and planes, creating what is called a ``blocking dip" in the flux of ions exiting from a thin enough crystal as a function of the angle with respect to a symmetry axis or plane. These effects  were first observed in the 1960's and are used in crystallography,  in the study  of lattice disorder, ion implantation, surfaces, interfaces and epitaxial layers, in measurements of short nuclear lifetimes etc.  In particular, avoiding channeling is essential in the manufacturing of semiconductor devices, since ion implantation at a controlled depth  is the primary technique.

Ion channeling in NaI (Tl) was first observed in 1973 by Altman \textit{et al.}~\cite{Altman}. They observed that channeled ions produce more scintillation light because they lose most of their energy via electronic stopping rather than nuclear stopping. The  potential importance of the channeling effect for direct  dark matter detection was first pointed out by  Sekiya \textit{et al.}~\cite{japanese}   and subsequently for NaI (Tl) by Drobyshevski~\cite{Drobyshevski:2007zj} and by the DAMA collaboration~\cite{Bernabei:2007hw}. When  Na or I ions recoiling after a collision with a  dark matter WIMP (Weakly Interacting Massive Particle) move along crystal axes and planes, their quenching factor is approximately $Q=1$ instead of $Q_{\rm Na}=0.3$ and $Q_{\rm I}=0.09$, since they give their energy to electrons.

Ion channeling in crystals could give rise to a daily modulation due to the preferred direction of the dark matter flux arriving on the Earth. Earth's daily rotation naturally changes the direction of the ``WIMP wind'' with respect to the crystal axes, which produces a daily modulation in the measured recoil energy (equivalent to a modulation of the factor $Q$). Sekiya \textit{et al.}~\cite{japanese} pointed this out for stilbene crystals, and Avignone, Creswick, and Nussinov~\cite{Avignone:2008cw} for NaI (Tl) crystals, although the available estimates of the strength of these daily modulations are somewhat simplistic. My collaborators, Graciela Gelmini and Paolo Gondolo, and I~\cite{BGG-I} present here our analytic calculations of the channeling fraction in NaI (Tl), Si and Ge crystals.

\section{Modeling of channeling}

Our calculation  is based on  the classical analytic models developed in the 1960's and 70's, in particular by Lindhard~\cite{Lindhard:1965, Andersen:1967, Morgan-VanVliet, VanVliet, Andersen-Feldman, Komaki:1970, Appleton-Foti:1977,Hobler}. We use the continuum string and plane model, in which the screened Thomas-Fermi potential is averaged over a direction
parallel to a row or a plane. This averaged potential $U$ is considered to be uniformly smeared along the row or plane of atoms, which is a good approximation if the propagating ion interacts with many lattice atoms in the row or plane by a correlated series of many consecutive glancing collisions with lattice atoms. Just one row or plane is considered in this model. Lindhard proved that for an ion propagating with kinetic energy $E$, and for small  angle $\phi$  between the ion's trajectory and  the atomic  row (or plane) in the direction perpendicular to  the row (or plane), the  so called ``transverse energy", $E_{\perp} = E \sin^2\phi + U\simeq E \phi^2 + U$ is conserved.

The conservation of the transverse energy provides a definition of the minimum distance of approach to the string (or plane) of atoms, $\rho_{\rm min}$, at which the trajectory of the ion makes  a zero angle with the string (or plane), and also of the angle $\psi$ at which the ion exits from the string (or plane), i.e. far away from it where $U \simeq0$. In reality the furthest position from   a string or plane of atoms is  the middle of the channel, whose width we call  $d_{\rm ch}$. Thus,
\begin{equation}
E_{\perp}= U(\rho_{\rm min}) =  E \psi^2 +U(d_{\rm ch}/2).
\label{eq:consetrans}
\end{equation}

Channeling requires that $\rho_{\rm min} > \rho_c(E)$, where $\rho_c(E)$ is the smallest possible minimum distance of approach of the propagating ion with the row (or plane) for a given energy $E$. Since the potential $U(\rho)$  decreases monotonically with increasing $\rho$, $U(\rho_{\rm min}) < U(\rho_c(E))$. Using Eq.~\ref{eq:consetrans}, this can be further translated into an upper bound on $E_{\perp}$ and  on  $\psi$, the angle the ion makes with the string far away from it,
\begin{equation}
\psi < \psi_{c}(E)=  \sqrt{ \frac{U(\rho_c(E))- U(d_{\rm ch}/2)}{E} }.
\label{psicritaxial}
\end{equation}

 $\psi_{c}(E)$  is the maximum angle  the ion can make with the string far  away from it (i.e. in the middle of the channel) if the ion is channeled. At low enough $E$,  $\rho_c(E)$ becomes close to $d_{\rm ch}/2$, and  the critical angle $\psi_{c}(E)$ goes to zero. This means that there is a minimum energy below which channeling cannot happen, even for ions moving initially in the middle of the channel.

\begin{figure}
  \includegraphics[height=150pt]{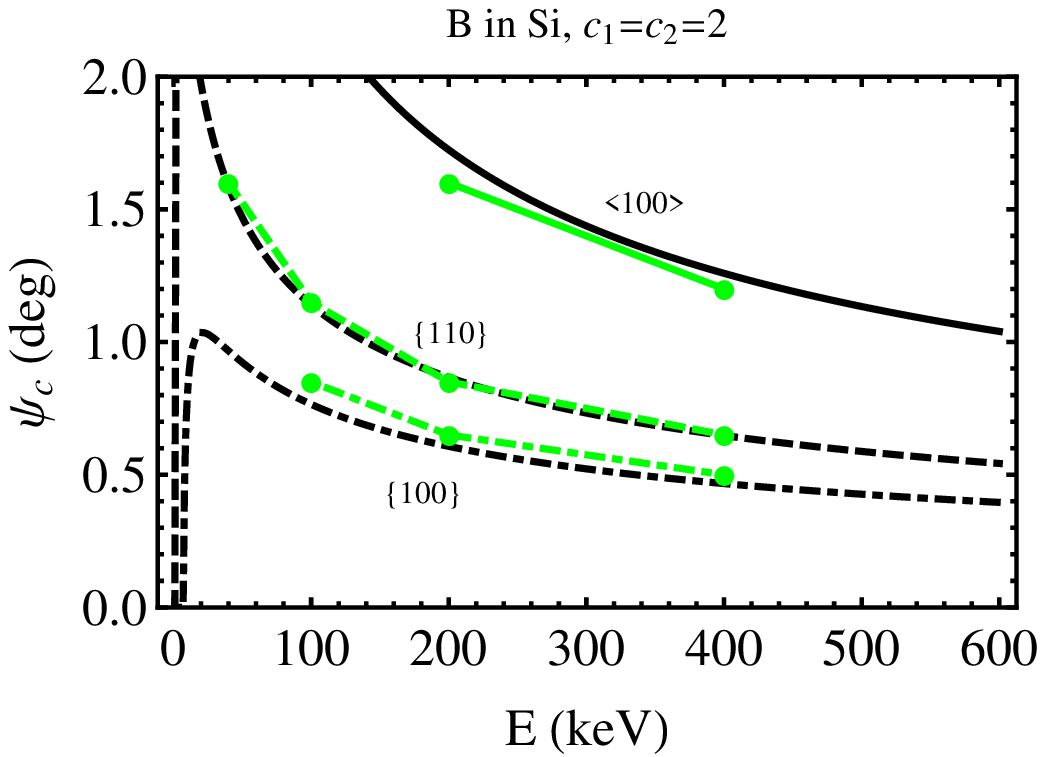}
  \includegraphics[height=155pt]{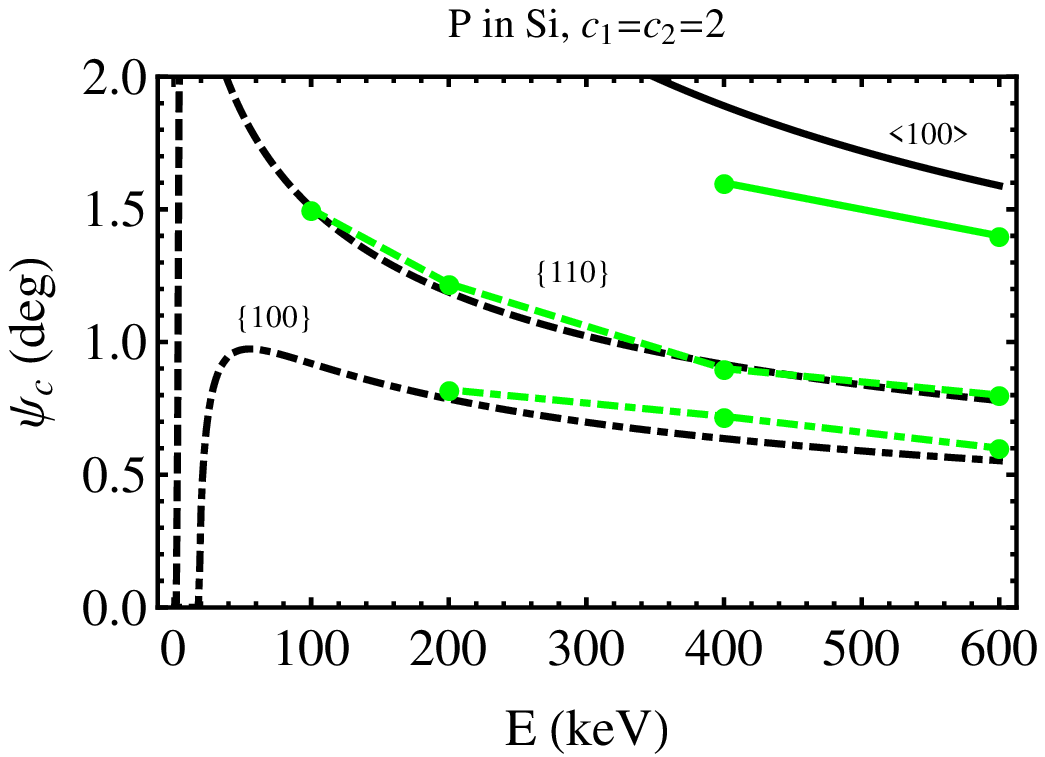}\\
  \caption{Comparison of theoretical (black lines) temperature corrected critical angles (with $c=2$) and measured critical angles extracted from thermal wave measurements~\cite{Hobler} (green, or gray if color not available, dots, joined by straight lines to guide the eye) as a function of the energy of (a) B  ions and (b) P ions propagating  in a Si crystal at T=20 $^{\circ}$C, for the indicated axial and planar channels.}
  \label{Compare-Data}
\end{figure}

So far we have been considering static strings and planes, but the atoms in a crystal are actually vibrating.  We use the Debye model, and take into account thermal effects  in the crystal through a modification of the critical distances which was  found originally by Morgan and Van Vliet~\cite{Morgan-VanVliet}  and later by Hobler~\cite{Hobler} to provide good agreement with simulations and data. It consists of taking the temperature corrected critical distance $\rho_c(T)$  to be,
\begin{equation}
\rho_c(T)= \sqrt{\rho^2_c(E) + [c u_1(T)]^2},
\label{rcofT}
\end{equation}
where $u_1$ is the one dimensional rms vibration amplitude  of the atoms in a crystal, and the factor $c$ in different references is a number between 1 and 2~\cite{Morgan-VanVliet, VanVliet}.
As shown in Fig.~\ref{Compare-Data}, with this formalism and using $c=2$ we  fit relatively well the critical angles   measured  at room temperature for  B and P ions  in a Si  crystal in several channels, for energies between 20 keV and 600 keV that Hobler~\cite{Hobler} extracted from thermal wave measurements.

\begin{figure}
 \includegraphics[height=170pt]{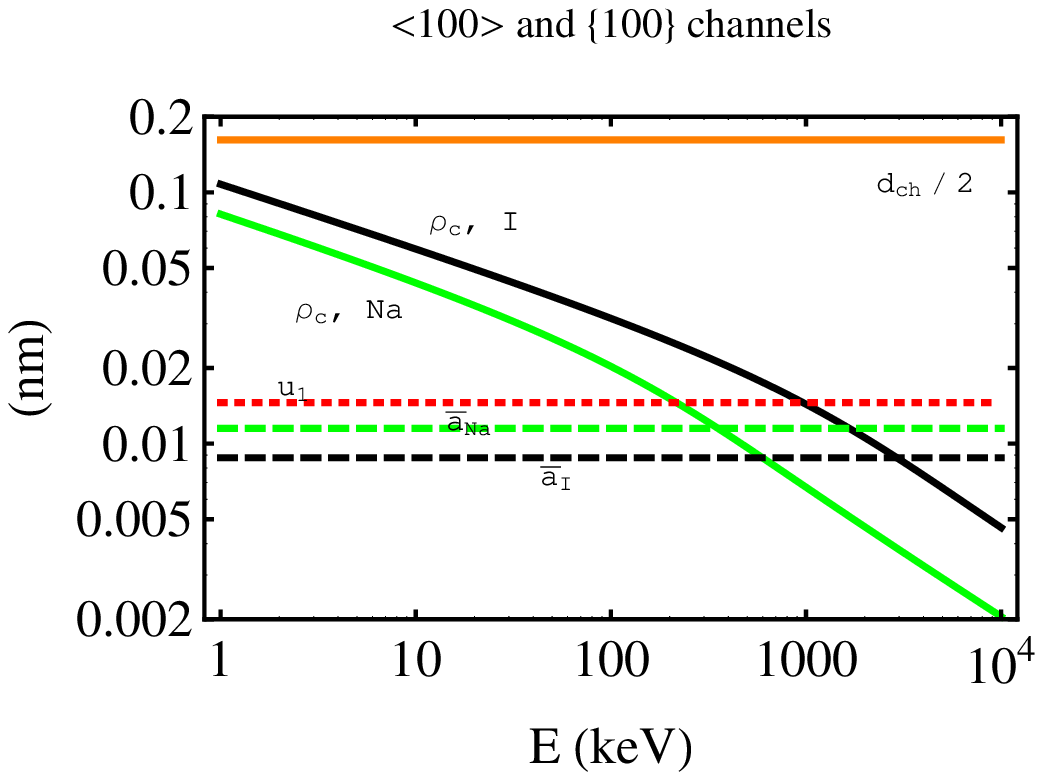}
 \includegraphics[height=170pt]{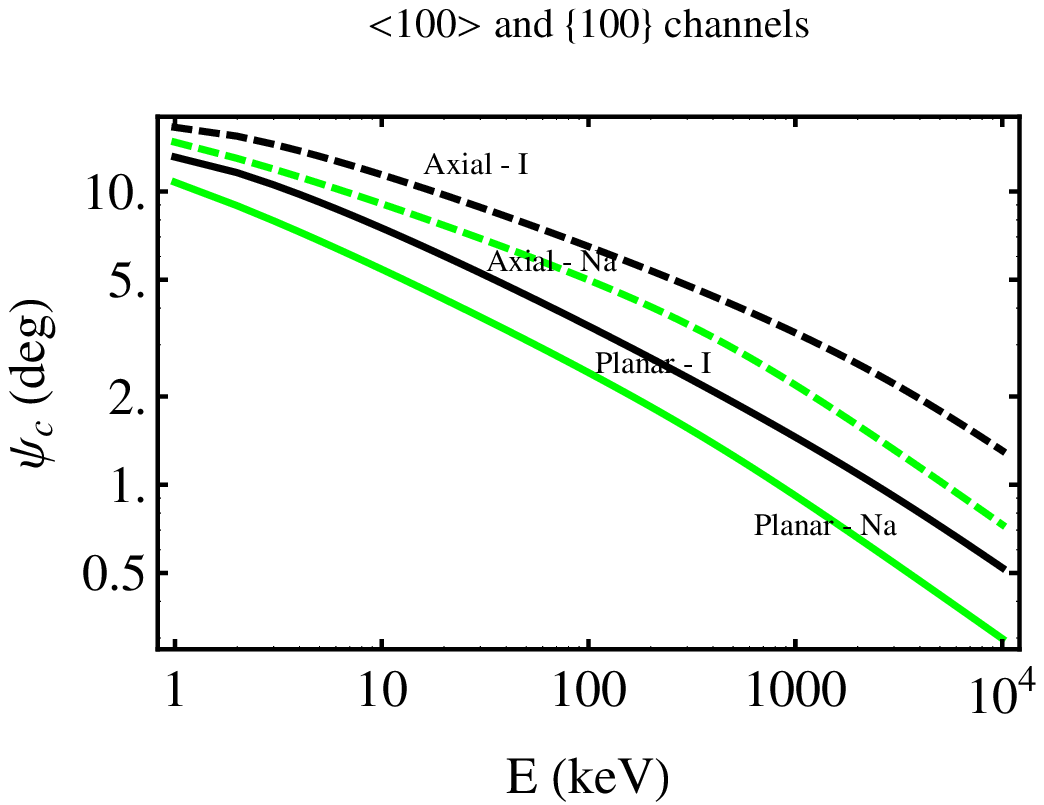}\\
  \caption{(a) Static critical distances of approach and  $u_1$ at 20 $^\circ$C and (b) critical channeling angles at 20 $^\circ$C with $c=1$ as a function of the energy of propagating Na (green/gray) and I (black) ions in the $<$100$>$ axial and \{100\} planar channels. The screening radii shown as vertical lines are $\bar{a}_{\rm Na}=0.00878$ nm and $\bar{a}_I=0.0115$ nm~\cite{BGG-I}.}
  \label{rc-psic-MV-100}
\end{figure}

As an example, the static ($c=0$) axial and planar critical  distances are presented in Fig.~\ref{rc-psic-MV-100}(a) for the 100 channel in NaI crystal, together with the amplitude of thermal vibration $u_1$ at 20 $^\circ$C, and the Thomas-Fermi screening distances for Na and I ions. Fig.~ \ref{rc-psic-MV-100}(b)  shows the  temperature corrected axial and planar  critical angles at 20 $^\circ$C  (with $c=1$) for the same channel as functions of energy of the traveling Na and I ions.

\section{Channeling of incident particles}

The channeling of ions in a crystal depends not only on the angle their initial trajectory makes with  strings or planes in the crystal, but also on their initial position. Ions which start their motion close to the center of a channel,  far from a string or plane, where they make an angle $\psi$, are channeled if the angle  is smaller than a critical angle (as explain earlier) and are not  channeled otherwise. Particles which start their motion in the middle of a channel (as opposed to  a lattice site) must be incident upon the crystal.

No data or simulations of Na and I ions propagating in a NaI crystal is available at low energies. We show that  to a good approximation we can use analytic calculations and reproduce the channeling fraction in NaI presented in Ref.~\cite{Bernabei:2007hw}. Temperature corrections are neglected and we use a static lattice (similar to the approach of Ref.~\cite{Bernabei:2007hw}). For an incident angle $\psi$ with respect to each of the channels and an ion energy $E$, the fraction ${\chi}_{\rm inc}(E,\psi)$ of channeled incident ions for axial and planar  channels  is ${\chi}_{\rm inc}=1$ if $\psi$ is smaller than the critical angle for the corresponding channel and zero otherwise.
\begin{figure}
   \includegraphics[height=190pt]{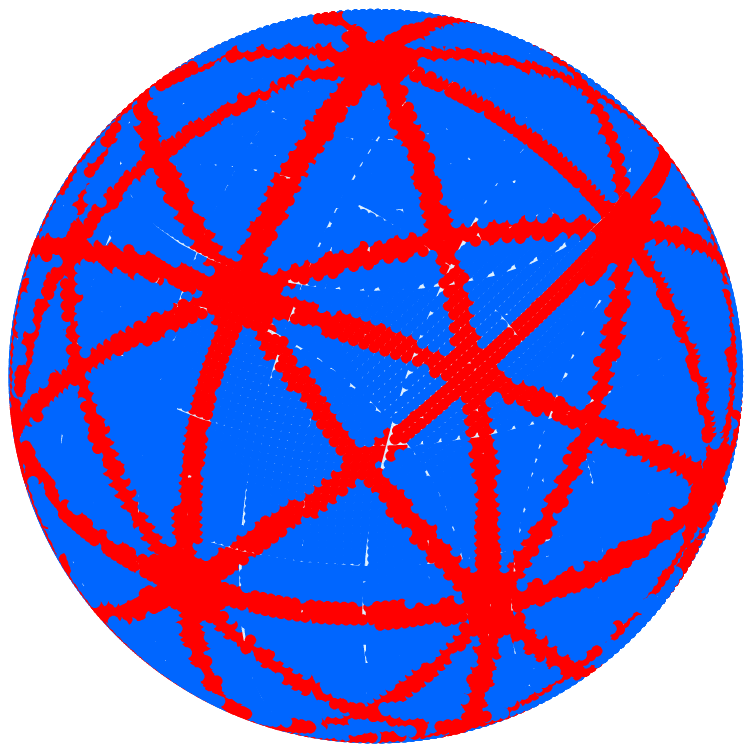}
  \includegraphics[height=160pt]{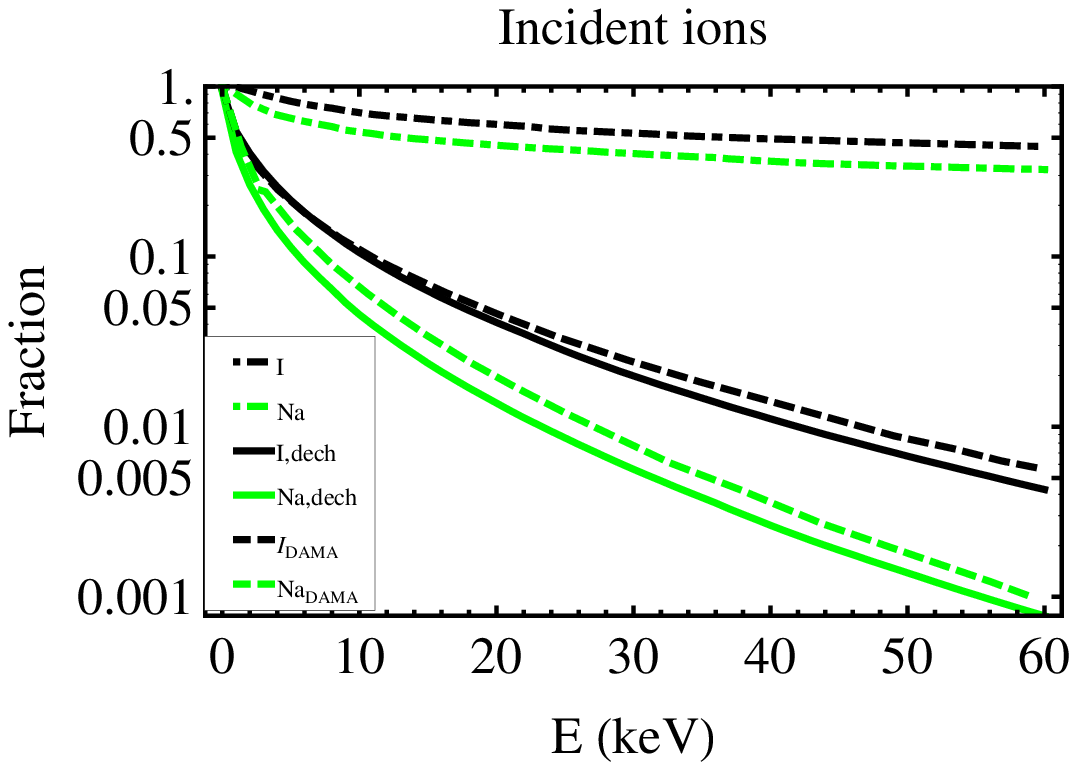}
 \caption{(a) Channeling fraction for a 50 keV Na ion in different directions plotted on a sphere using the HEALPix pixelization:  probability equal to one in red, and probability equal to zero in blue. (b) Fraction of channeled incident  I (black) and Na (green/gray) ions as a function of their incident energy $E$ with the static lattice without  (dot dashed lines) and with (solid lines) dechanneling due to interactions with Tl impurities.  The results of DAMA are also included (dashed lines).}
  \label{DAMA}
\end{figure}

To find the total fraction of channeled incident nuclei, we average ${\chi}_{\rm inc}$ over  the incident direction. The integrals cannot be solved analytically, so we integrated numerically by performing a Riemann sum once the sphere of directions has been divided using the Hierarchical Equal Area iso-Latitude Pixelization (HEALPix) method~\cite{HEALPix:2005}. The HEALPix method uses an algorithm to deal with the pixelization of data on a sphere, and it is useful to compute integrals over direction by dividing the surface of a sphere into many pixels, computing the integrand at each pixel (i.e. each direction, see Fig.~\ref{DAMA}(a)), and finally summing up the values of all pixels over the sphere.

A  channeled ion can be pushed out of a channel by  an interaction with an  impurity such as the atoms of Tl in NaI (Tl). Here we will simply assume that if a channeled ion interacts with a Tl atom it becomes dechanneled and thus it does not contribute to the fully channeled fraction any longer. We thus neglect  the possibility that after the interaction the ion may reenter into a channel, either the same or another. Fig.~\ref{DAMA}(a) shows the axial and planar channels of the NaI crystal for  incoming  Na ions with an energy of 50 keV. We include here only the channels with lower crystallographic indices 100, 110 and 111.  The fraction of channeled incoming Na  and I ions after  including  dechanneling in the way just described  is shown in  Fig.~\ref{DAMA}(b). Our results agree well with those published by the DAMA collaboration~\cite{Bernabei:2007hw}, also included in the figure.

\section{Channeling of recoiling lattice nuclei}

The recoiling nuclei start initially from lattice sites (or very close to them), thus blocking effects  are important. In fact, as argued originally by Lindhard~\cite{Lindhard:1965}, in a perfect lattice and in the absence of energy-loss processes the probability that a particle starting from a lattice site is channeled would be zero. The argument uses statistical mechanics in which the probability of particle paths related by time-reversal is the same. Thus the probability of an incoming ion to have a particular path within the crystal is the same as the probability of the same ion to move backwards along the same path~\cite{Gemmell:1974ub}. This is what Lindhard called the ``Rule of Reversibility". Using this rule, since the probability of an incoming channeled ion to get very close to a lattice site is zero, the probability of the same ion to move in the time-reversed path,  starting at a nuclear site and ending inside a channel, is zero too. However, any departure of  the actual lattice from a perfect lattice, for example due to vibrations of the atoms in the lattice, would violate the conditions of this argument and allow for some of the recoiling lattice nuclei to be channeled, as already understood in the 70's~\cite{Komaki:1970, Komaki-et-al-1971}. We now estimate the effect using the formalism presented so far.

For axial channels, the probability distribution function $g(r)$ of the perpendicular distance to the row of the colliding atom due to thermal vibrations can be represented by a two-dimensional Gaussian. The channeled fraction of nuclei with recoil energy $E$  making an initial  angle $\phi$ with respect to the axis is given by the fraction of nuclei which can be found at a distance $r$ larger than  minimum distance $r_{i,\rm min}$ from the row at the moment of collision,
\begin{equation}
\chi_{\rm axial}(E, \phi)=\int_{r_{i,\rm min}}^{\infty}{dr g(r)}=\exp{(-r_{i,\rm min}^2/2u_1^2)}.
\label{chiaxial}
\end{equation}

\begin{figure}
  \includegraphics[height=160pt]{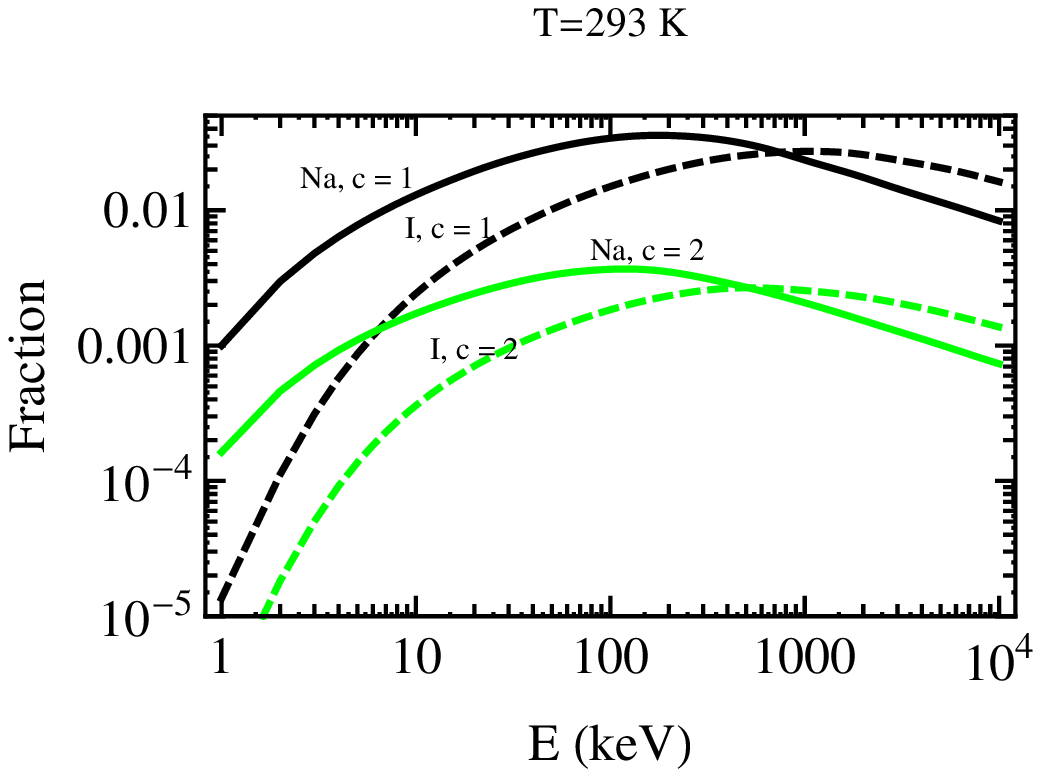}
  \includegraphics[height=160pt]{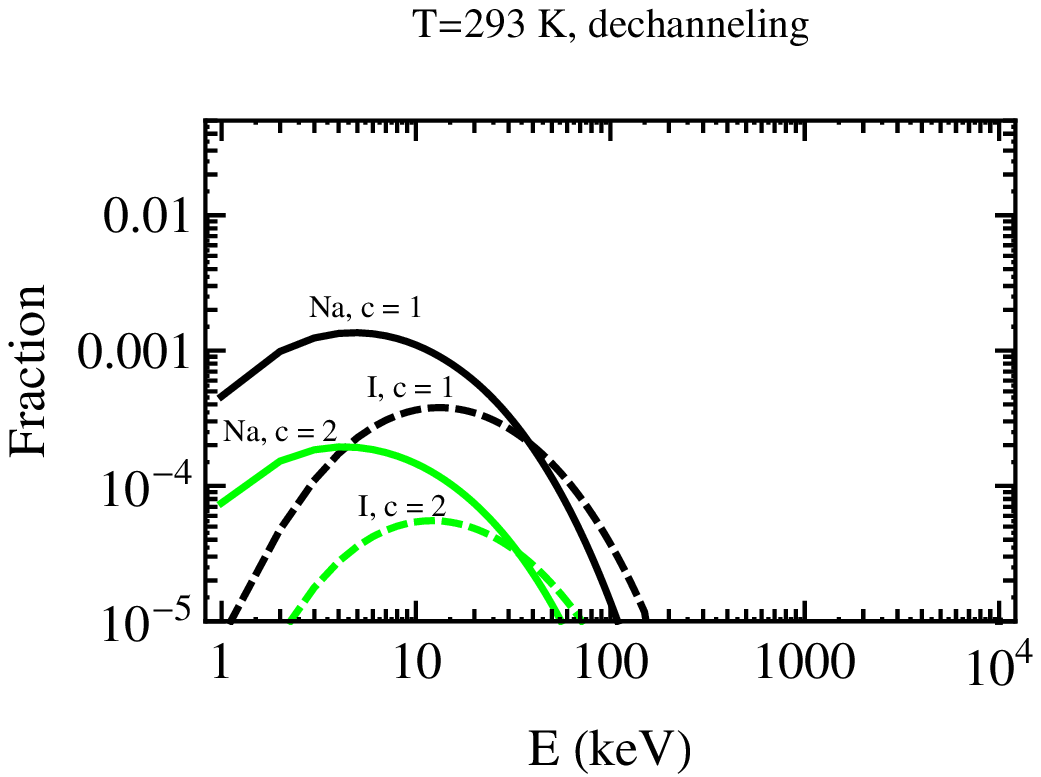}\\
  \caption{Channeling fractions at T=293 K for Na (solid lines) and I (dashed lines) ions for $c=1$ (black) and $c=2$ (green/gray) cases (a) without and (b) with dechanneling included.}
  \label{FracNaI-Final}
\end{figure}

Figure \ref{FracNaI-Final} shows what we consider to be our main predictions for the range expected as an upper limit to the channeling fraction in NaI (Tl) at 239 K for two different assumptions for the effect of thermal vibrations in the lattice, which depend on the value
of the parameter $c$ used in  the temperature corrected critical distances of approach. Dechanneling is ignored in Fig.~\ref{FracNaI-Final}(a) and taken into account in Fig.~\ref{FracNaI-Final}(b).

\begin{figure}
  \includegraphics[height=160pt]{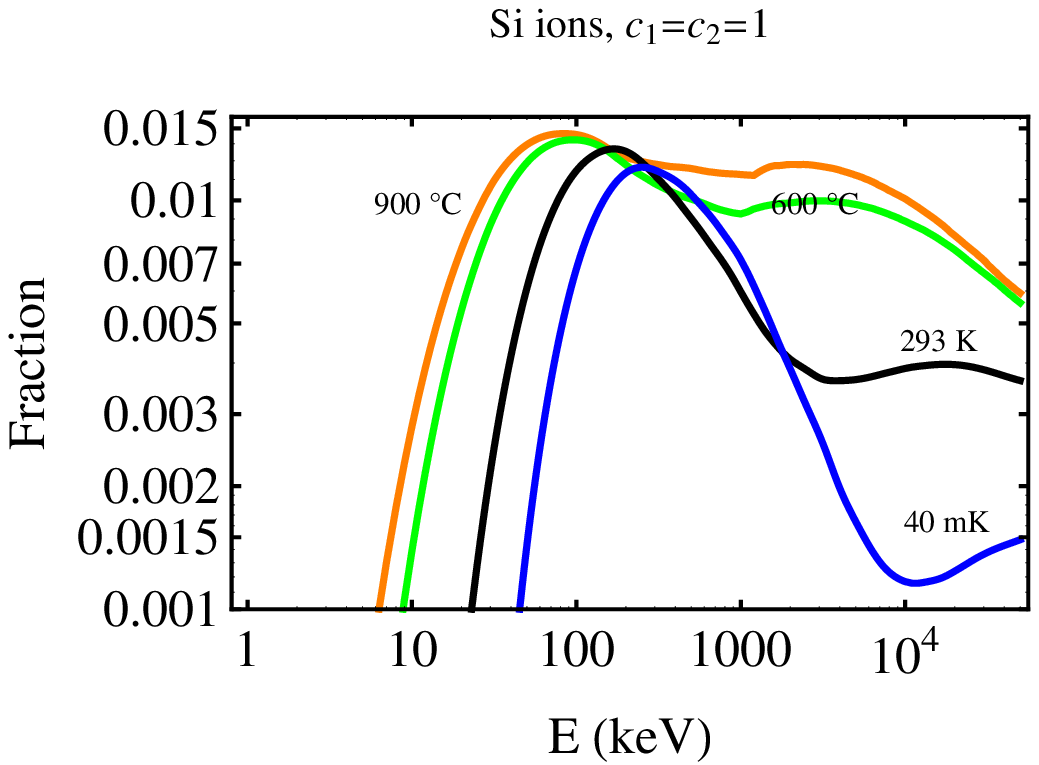}
  \includegraphics[height=160pt]{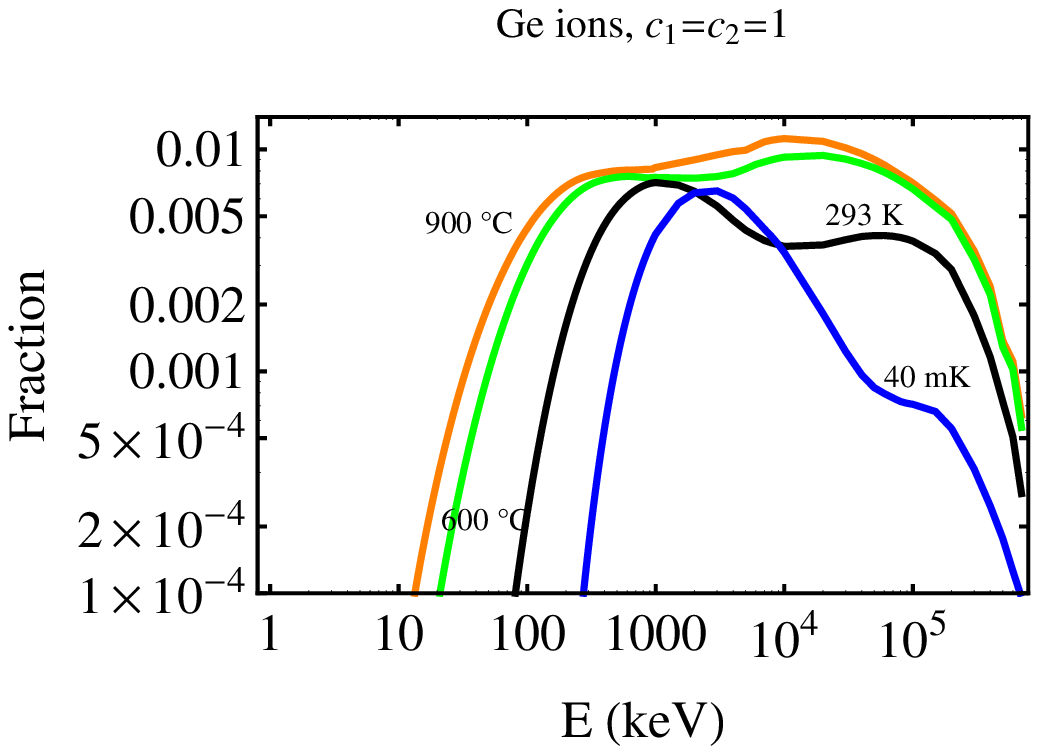}\\
  \caption{Channeling fractions of (a) Si  and (b) Ge recoils in a Si and a Ge crystal respectively, as a function of the ion energy for  temperatures T=900 $^\circ$C (orange or medium gray), T=600 $^\circ$C (green or light gray), 293 K (black), and 44 mK (blue or dark gray)  in the approximation of $c=1$.}
  \label{FracSiG-DiffT-c1}
\end{figure}

\begin{figure}
  \includegraphics[height=160pt]{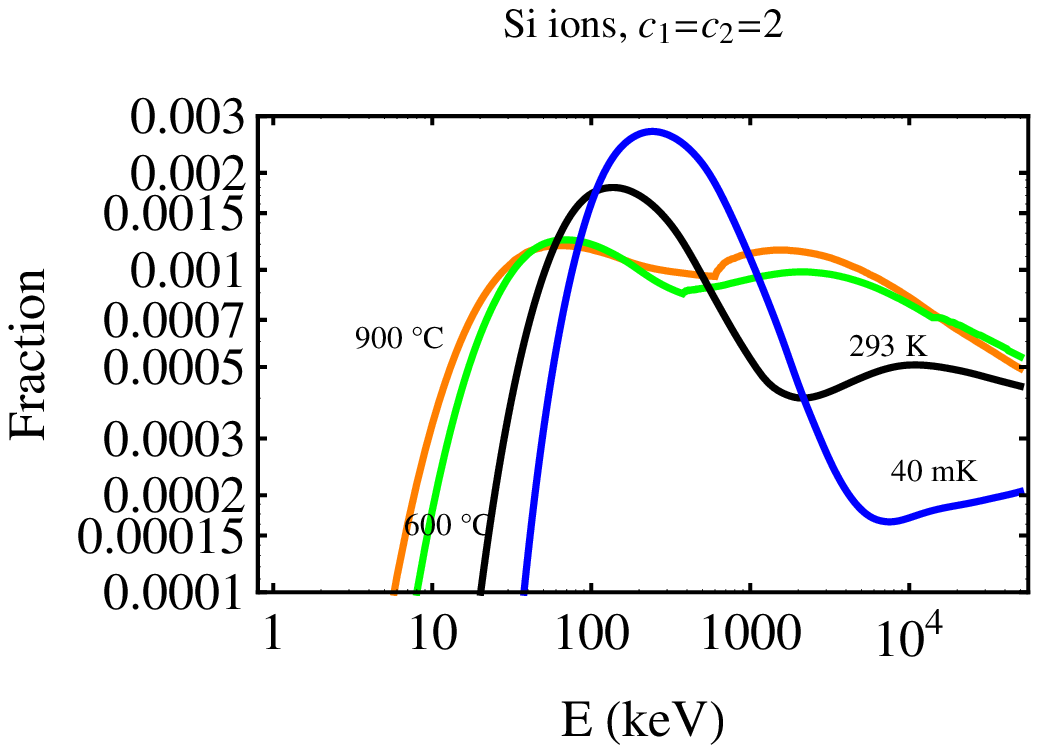}
  \includegraphics[height=160pt]{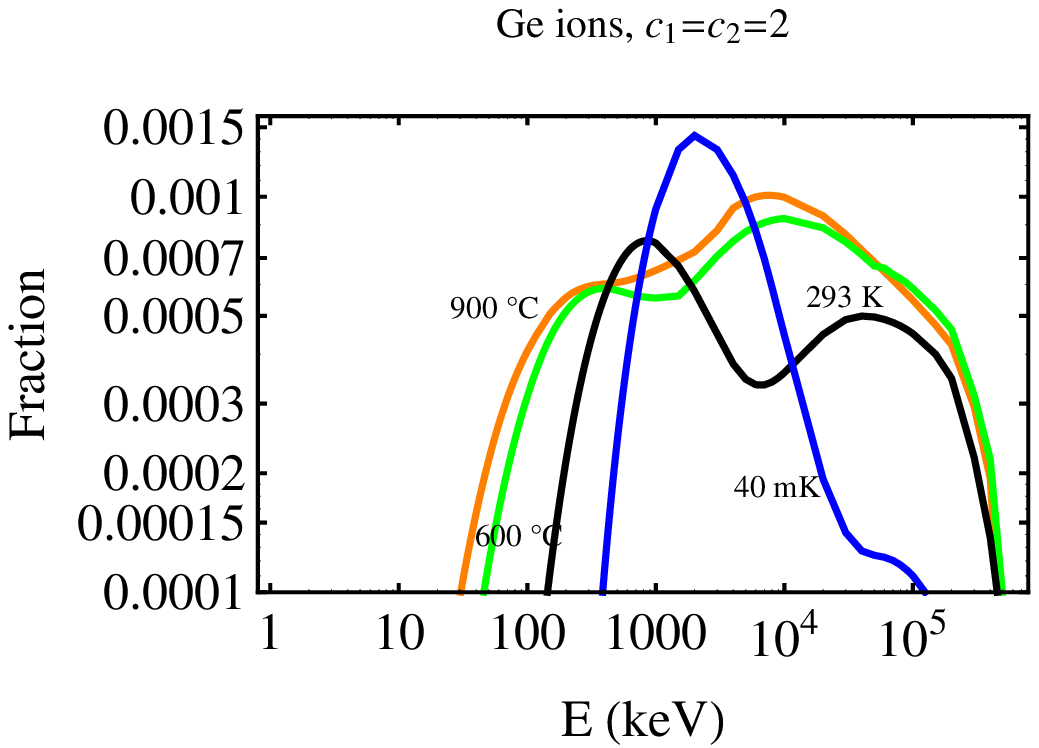}\\
  \caption{Same as Fig.~\ref{FracSiG-DiffT-c1} but for $c=2$.}
  \label{FracSiG-DiffT-c2}
\end{figure}

Our results for the geometric total channeling fraction for Si ions propagating in a Si crystal  and  Ge ions propagating in a Ge crystal at different temperatures are shown in Figs.~\ref{FracSiG-DiffT-c1} and \ref{FracSiG-DiffT-c2} for the two cases of $c=1$ and $c=2$, respectively. Please note that we have not considered the possibility of dechanneling of initially channeled ions, due to imperfections in Si and Ge crystals. Any mechanism of dechanneling will decrease the fractions obtained here for Si and Ge.

\section{Conclusions}

We have studied the channeling of ions recoiling after collisions with WIMPs within NaI (Tl), Si and Ge crystals.   Channeled ions move within the crystal along symmetry axes and planes and suffer a series of small-angle scatterings  that maintain them in the open ``channels"  between the rows or planes of lattice atoms and thus penetrate much further into the crystal than in other directions. Ions which start their motion close to the center of a channel, at an initial angle $\psi$, are channeled if the initial angle  is smaller than the critical angle in Eq.~\ref{psicritaxial}, and are not  channeled otherwise. We have found that the channeling of lattice ions recoiling after a collision with a WIMP is very different from the case of incident ions, and that the channeling fraction is smaller.

As argued originally by Lindhard~\cite{Lindhard:1965}, in a perfect lattice and in the absence of energy-loss processes, the probability that a particle starting from a lattice site is channeled would be zero. However, due to vibrations  in the crystal, the atom that interacts with a WIMP may be displaced from its position in a perfect lattice, and there is a non-zero probability of channeling (as given in Eq.~\ref{chiaxial}).

 As we see in Fig.~\ref{FracNaI-Final}(a) for NaI, without including dechanneling, the channeling fraction is never larger than 5\% and the maximum happens at 100's of keV. This maximum occurs because the critical distances decrease with the ion energy $E$, making channeling more probable, and the critical angles also decrease with $E$, making channeling less probable. The simple extreme model of dechanneling we used for NaI predicts much smaller fractions, at most in  the 0.1\% level, with the maximum shifted to small energies, less than 10 keV (see Fig.~\ref{FracNaI-Final}(b)). This reduction may eventually prove to be too extreme and at present we do not have a better formalism to model dechanneling. With the simple model of dechanneling we used for NaI, we could reproduce the channeling fractions computed by the DAMA collaboration (which, however, apply to ions which start their motion close to the middle of a channel and not to the case of WIMP direct detection).

If the values found by Hobler~\cite{Hobler}, and also by us above (see Fig.~\ref{Compare-Data}), to reproduce measured channeling angles in B and P propagating in Si apply also to the propagation of Si ions in Si, then the case of
 $c=2$ should be chosen and the channeling fractions would never be larger than 0.3\%. Moreover, increasing the temperature of a crystal usually increases the fraction of channeled recoiling ions (see Fig.~\ref{FracSiG-DiffT-c1}), but not always. Sometimes  the opposite happens (see Fig.~\ref{FracSiG-DiffT-c2}). The $c=1$ choice leads to channeling fractions close to 1\% for Si and Ge.

 The analytical  approach used here can successfully describe qualitative features of the channeling and blocking effects, but should be complemented by data fitting of parameters and by simulations to obtain a good quantitative description too.  Thus our results should in the last instance be checked by using some of the many sophisticated Monte Carlo simulation programs.

\section{acknowledgments}

N.B. and Graciela Gelmini were supported in part by the US Department of Energy Grant
DE-FG03-91ER40662, Task C. Paolo Gondolo was supported in part by  the NFS
grant PHY-0456825 at the University of Utah. We would like to thank S. Nussinov and F. Avignone for
several important discussions about their work, and  J. U Andersen, D. S.  Gemmell, D. V. Morgan, G. Hobler, and Kai Nordlund for some exchange of information.
We also thank P. Belli for several conversations.

\end{document}